\documentclass[aps,prl,showpacs,twocolumn,superscriptaddress]{revtex4-1}
\usepackage{graphicx,subfigure}

\usepackage{amsmath}
\usepackage{bm}
\usepackage{graphics}
\usepackage{dsfont}
\usepackage{epsfig}

\renewcommand{\Re}{ \operatorname{Re} }
\renewcommand{\Im}{ \operatorname{Im} }

\begin{document}
\title{Fractional Spin Josephson Effect and Electrically Controlled Magnetization in Quantum Spin Hall Edges}

\author{Qinglei Meng}
\affiliation{Department of Physics, University of Illinois, Urbana, IL 61801}

\author{Vasudha Shivamoggi}
\affiliation{Department of Physics, University of Illinois, Urbana, IL 61801}
\affiliation{Department of Electrical and Computer Engineering, University of Illinois, Urbana, IL 61801}

\author{Taylor L. Hughes}
\affiliation{Department of Physics, University of Illinois, Urbana, IL 61801}

\author{Matthew J. Gilbert}
\affiliation{Department of Electrical and Computer Engineering, University of Illinois, Urbana, IL 61801}
\affiliation{Micro and Nanotechnology Laboratory, University of Illinois, Urbana, IL 61801}

\author{Smitha Vishveshwara}
\affiliation{Department of Physics, University of Illinois, Urbana, IL 61801}

\date{\today}

\begin{abstract}
We explore a spin Josephson effect in a system of two ferromagnets coupled by a tunnel junction formed of 2D time-reversal invariant topological insulators. In analogy with the more commonly studied instance of the Josephson effect for charge in superconductors,  we investigate properties of the phase-coherent {\it spin} current resulting from the misalignment of the in-plane magnetization angles of the two ferromagnets. We show that the topological insulating barrier offers the exciting prospect of hosting a {\it fractional} spin Josephson effect mediated by bound states at the ferromagnet-topological insulator interface. We provide multiple perspectives to understand the $4\pi$ periodic nature of this effect. We discuss several measurable consequences, such as, the generation of a transverse voltage signal which allows for purely electrical measurements,  an inverse of this effect where an applied voltage gives rise to a transverse spin-current, and a fractional AC spin-Josephson effect.
\end{abstract}


\maketitle

\begin{figure}[t]
\begin{center}
\includegraphics[width=0.3\textwidth]{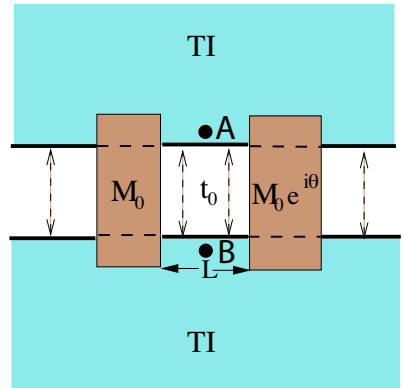}
\end{center}
\caption{\label{twoedge}  Two ferromagnetic islands connected by two-edges of a 2D topological insulator which are themselves tunnel coupled. Bound states mediate a fractional spin Josephson effect which gives rise to a $4\pi$ periodic spin current and voltage signal between points A and B as a function of the winding of the relative phase between the ferromagnets.}
\end{figure}

The recent discovery of 2D time-reversal invariant topological insulators (TIs)\cite{kane2005A,bhz2006,koenig2007} has generated a wide-ranging collection of device proposals. The interplay between spin-orbit coupling and magnetic\cite{qhz2008,spinbatterypaper,foldingpaper} or superconducting\cite{kanefu3d,FuKaneJJ,beenakker} islands proximity coupled to the TI edge states has led to the prediction of localized fractional charge\cite{qhz2008,foldingpaper}, quantized pumping of electrical current\cite{qhz2008,spinbatterypaper}, Majorana bound states\cite{kanefu3d,FuKaneJJ,beenakker}, and a fractional Josephson effect\cite{FuKaneJJ,beenakker}.We expand these ideas by focusing on the generation of spin currents in magnetic-coupled TI edge states.  In this letter,
we propose coupling a ferromagnetic junction to the 2D TI edge states, as shown in Fig.~\ref{twoedge}, to produce a fractional \emph{spin}-Josephson current with a $4\pi$-periodicity. We exploit an analogy between spin in a ferromagnet and charge in a superconductor to explain this unconventional transport phenomenon as the magnetic analog of the fractional (charge) Josephson effect mediated by Majorana bound states~\cite{FuKaneJJ}. While spin effects are naturally present in TI devices due to the strong spin-orbit coupling, we discuss purely electrical signatures of this spin effect which would be experimentally accessible via quantum transport.

We briefly review  standard superconducting (SC) Josephson junction physics to set the stage for the spin Josephson effect (SJE) analog.  An S-I-S junction consists of two SC regions separated by an insulating barrier.  At zero bias-voltage, while the SC gap prevents a single electron from tunneling, charge current can result from  Cooper pairs tunneling across the barrier at zero energy cost.  The phase difference between the SC order parameters on the left and right sides of the junction, $\Delta\phi =\phi_R-\phi_L$, determines the properties of this Josephson current, and  is canonically conjugate to the \emph{difference} in the number of Cooper pairs $N=N_R-N_L$:
\begin{equation}
[\Delta\phi, N] = i,
\label{SCcommutation}
\end{equation}
  The form of the current is determined by the Hamiltonian for the junction:
\begin{equation}
H_{SC} =-E_J \cos (\Delta \phi) +{\small{\frac{2e^2}{C}}}N^2
\label{JJham}
\end{equation}\noindent where $C$ is the capacitance of the tunnel junction, and $E_J>0,$.
Specifically, the Josephson current  $I = 2e\langle \dot{N}\rangle = -2e i[N, H_{SC}]/ \hbar = -2e E_J \sin \Delta\phi / \hbar$ is driven by a difference in the phases of the order parameters rather than an applied voltage, making it a dissipationless supercurrent. In the presence of an applied voltage $V=2eN/C$ the equation of motion for the phase is $\Delta\dot{\phi}=2eV/\hbar$, yielding the AC Josephson effect.

In fact, a phase-induced Josephson-like current can arise in a variety of systems having phase coherence, where the ``charge" is the appropriate quantity that is canonically conjugate to the  phase difference.  One notable example is in quantum Hall bilayers, where phase coherence between the layers has been used to explain a zero-bias conductance peak~\cite{spielman2000}.  Here, we focus on the case of a tunnel junction between two ferromagnetic (FM) insulators, first establishing the analogy with the standard superconductor Josephson physics.   Phase-coherent tunneling between two FMs across a non-magnetic barrier can thus produce a spin current analogous to the charge current in the SC case. Such a SJE  has been observed in He$^3$ thin films~\cite{BorovikHe3}, and proposed to exist in a FM junction having an excitonic insulator barrier~\cite{WangSpinlike}. 
Josephson-like physics requires a magnetic easy-plane anisotropy, either intrinsic  or induced by a substrate material (as in \emph{e.g.} Ref. \onlinecite{NogueiraBennemann}), giving rise to an effective ``spin"-capacitance. 
Each ferromagnet is characterized by an in-plane order parameter $M_0 e^{i\theta_{L/R}}$ (right/left FMs). The phase angles  $\theta_{L/R}$, which define the directions of the magnetization in the easy-plane, are canonically conjugate to the  z-component of the total spin in each of the FMs (denoted $S^z_{L/R}$)\cite{NogueiraBennemann, Villain}: 
\begin{equation}
[\theta_{L/R} , S^z_{L/R}] = i\hbar.
\label{FMcommutation}
\end{equation}
To explicitly see how the conjugate relationship results in phase coherent spin current, consider an FM tunnel junction connecting regions with unequal phases, $\theta _L\neq\theta_R.$ The FM junction can be described by the Hamiltonian
\begin{equation}
H_{FM} =-E_S \cos (\Delta\theta) + \alpha(S_{L}^{z})^2+ \alpha(S_{R}^{z})^2
\label{SJJham}
\end{equation}\noindent where $\Delta\theta=\theta_R-\theta_L$, $E_s$ reflects the exchange coupling between the ferromagnets and the terms proportional to $\alpha$ represent the magnetic-anisotropy induced  `spin-capacitance' .
 The spin current across the junction, $I_s = \langle dS^z _R/dt \rangle =-i[S^z _R, H_{FM}]/\hbar$, becomes $I_S = -E_S \sin \Delta\theta.$ Additionally the rate of change of phase  $\Delta\dot{\theta}=-i[\Delta\theta,H_{FM}]/\hbar=2\alpha(S_{R}^z-S_{L}^z)$ can be compared to $\Delta\dot{\phi}=2eV/\hbar$ for the AC charge Josephson effect.

The analogy between the charge current in the SC case and spin current in the FM case may be made explicit by performing a particle-hole conjugation on one of the spin sectors~\cite{AndersonPseudo}. In the former case, a spin-up electron impinging the SC cannot be transmitted through but can Andreev reflect as a spin-down hole (with angular momentum $+\hbar/2$ assuming $s$-wave pairing in the SC). The net effect is to transport a Cooper pair of charge $2e$ and zero spin into the superconductor.  An analogy was proposed in Ref.~\onlinecite{WangSpinlike} to explain a similar effect in a FM junction between two excitonic insulators.  The FM regions impose an energy cost to a single spin-up electron.  The FM regions can be heuristically described as polarized \emph{excitonic} condensates themselves which can absorb a ferromagnetic exciton, consisting of a spin-up electron and a spin-down hole, at zero energy.  Therefore a spin-up electron incident on an FM region can be reflected back as a spin-down electron, as the FM region absorbs an exciton pair and $\hbar$ spin.  In this case, no net charge is transported into the FM region, but there is a non-zero spin current.   

Turning to the fractional Josephson effect,  it is once again instructive to first review the SC case discussed in Ref. \onlinecite{FuKaneJJ}.  Consider a Josephson junction comprised of two $s$-wave superconductors separated by a FM barrier, all on a single TI edge.  The proximity-coupling to the SC and FM regions opens a gap in the edge states but the system supports mid-gap modes, one localized at each end of the junction at the places where the two competing mass terms are equal\cite{FuKaneJJ}.  These bound states are Majorana fermions, quasiparticles that are their own antiparticle.  The presence of these states alters the transport properties of the Josephson junction as the  Majorana bound states mediate the transfer of  \textit{single} electrons, as opposed to Cooper pairs, across the junction~\cite{KitaevQW, KwonSenguptaYakovenko, FuKaneJJ}.  The resulting Josephson current goes as $I \propto \sin \Delta\phi /2$ and is thus $4\pi$-periodic in the phase difference, in contrast to the $2\pi$-periodic expression found in typical Josephson junctions.  

The question we examine here is how to create a fractional \textit{spin}-Josephson effect using an analogous FM junction. Following the arguments of Ref. \onlinecite{foldingpaper}, we find that the relevant mass term that competes with the FM mass gap is an inter-edge tunneling term. 
Thus, we consider a junction consisting of two FM regions coupled to the edge states of two TI systems (Fig.~\ref{twoedge}). An alternative would be a single 2D TI with an etched or ablated weak link that would serve as the tunnel junction region.
Either way we assume that electrons can tunnel between the lower edge of one and upper edge of the other with tunneling amplitude $t_0$ and interact with the magnetic order parameter on the islands via the Zeeman coupling\cite{spinbatterypaper}.  In the basis $\left( c_{\uparrow, top} ~c_{\downarrow, top} ~c_{\uparrow, bot} ~c_{\downarrow, bot}\right) ^T$, the Hamiltonian is
\begin{equation}
H = -i\hbar v\partial _x \tau^z \sigma^z + \Re M(x) \sigma ^x + \Im M(x) \sigma ^y + t(x) \tau ^x ,
\label{QSHedgeHam}
\end{equation}
where $M(x), t(x)$ represent spatial dependent magnetic and tunneling terms respectively, $\sigma ^i (\tau ^i)$ are Pauli matrices acting on the spin (edge) sector and the tensor product is implicit. We note that Eq.~\ref{QSHedgeHam} has the same matrix structure, up to a unitary transformation, to that of the Josephson junction on the 2D TI edge discussed above\cite{FuKaneJJ}. The essential difference is the identification of the real and imaginary parts of the SC order with the $x$ and $y$ components of the in-plane magnetization (as expected from the charge/spin analogy) and the replacement of the competing magnetic gap in the SC case with the competing tunnel gap in the FM case.

Consider the magnetization in the FM regions lying in the plane perpendicular to the spin polarization of the TI edge states (for example, for the Bernevig-Hughes-Zhang model of Ref. \onlinecite{bhz2006} the magnetization would lie in the plane of the TI system). This is important because a magnetization in the same direction as the TI spin-polarization will not open a gap. 
The inter-edge tunneling and Zeeman coupling open competing gaps in the TI edge spectrum; for a uniform system, the gap is equal to the minimum of $|t_0\pm M_0|$.  For the junction geometry shown in Fig.~\ref{twoedge}, $M(x)$ vanishes inside the junction and the gap saturates to $t_0$. In the proximity of one of the magnets we have $|M_0| > |t_0|$. The  energy gap thus switches sign leading to a  trapped mid-gap electron state on each end of the junction where $|t_0|=|M_0|.$  An analytic solution can be obtained when the inter-edge tunneling is restricted to the region between the FM  islands, described by the mass profile

\begin{eqnarray}
M(x)& =& M_0 \Theta (-x) + M_0 e^{i\theta} (x-L),
\label{MagParam}\\
t(x) &=& t_0 \Theta (-x + L) \Theta (x) .
\label{HopParam}
\end{eqnarray}

\begin{figure}[t]
\begin{center}
\includegraphics[width=0.48\textwidth]{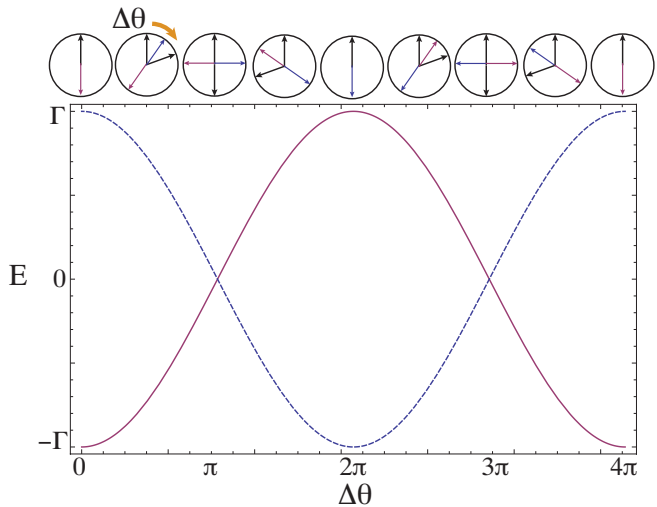}
\end{center}
\caption{\label{fig:energySJE} Energy of two boundstates as a function of $\Delta\theta.$, where $\Gamma=\frac{2M_0t_0 e^{-L t_0/\hbar v}}{M_0+t_0}.$ The top portion of figure indicates the in-plane spin of the bound-states (red(solid)/blue (dashed) color coded) and the magnetization directions of the two ferromagnets (black arrows). The system only returns to the original state when $\Delta\theta\to\Delta\theta+4\pi.$}
\end{figure}

In this case, there are two bound states, $b _L$ and $b _R$, localized at $x = 0$ and $x = L$ respectively, which are coupled through the effective Hamiltonian 
\begin{eqnarray}
H(\Delta\theta)&=&\Gamma\cos\frac{\Delta\theta}{2}{\textbf{b}}^{\dagger}\mu^y {\textbf{b}}^{\phantom{\dagger}} \equiv F(\Delta\theta){\textbf{b}}^{\dagger}\mu^y {\textbf{b}}^{\phantom{\dagger}}\label{eq:Ftheta}\\
\Gamma&=&\frac{2M_0t_0 e^{-L t_0/\hbar v}}{M_0+t_0}\nonumber
\end{eqnarray}\noindent where ${\textbf{b}}=(b_{L}\; b_{R})^{T}$ and $\mu^y$ is  Pauli matrix for the bound-state subspace and not related to physical spin. The energies of the boundstates are $E_{\pm}=\pm F(\Delta\theta)$ and are plotted in Fig. \ref{fig:energySJE}. It is important to observe that, as written, it appears that  $H(\Delta\theta+2\pi)=-H(\Delta\theta)$ but this is because we have performed a gauge transformation on the $b_{L/R}$ so that if, for example $\theta_{L}$ is fixed and $\theta_{R}$ advances by $2\pi$ then ${\bf b}\to \mu^z {\bf b}$ and $H(\Delta\theta)$ remains invariant.
We see that at $\Delta\theta=(2n+1)\pi$ there exist degeneracies in the spectrum at $E=0.$ 
Assuming that the other occupied modes do not contribute, the spin current is obtained from the derivative of $E_{\pm}(\Delta\theta)$
\begin{equation}
I_s (\Delta\theta) = \pm \frac{1}{2}\Gamma \sin \frac{\Delta\theta}{2}.
\label{SpinCurrent}
\end{equation}
Eq.~\ref{SpinCurrent} is the magnetic analog of the result in Ref.~\onlinecite{FuKaneJJ}: a gradient in the phase of the magnetic order parameter drives a spin current across the junction.  As the magnetization at the right end of the junction rotates by $2\pi$, the Hamiltonian returns to its original form,  while $I_s(\Delta\theta) \neq I_s (\Delta\theta + 2\pi).$ This indicates that the system experiences a non-trivial change when $\Delta\theta\to\Delta\theta+2\pi,$  only returning to its initial state after $\Delta\theta\to\Delta\theta+4\pi.$ A simple physical picture illustrates the nature of this periodicity (see top portion of Fig. \ref{fig:energySJE}). The average spin-polarization ($\vec{S}$) of the bound states is 
\begin{eqnarray}
\vec{S}&=&\pm\frac{\Xi}{2}\left(\cos{\tfrac{\theta_{L}+\theta_{R}}{2}},\sin\tfrac{\theta_{L}+\theta_{R}}{2},0\right)\\
\Xi&=&2M_0 t_0\frac{\frac{L}{\hbar v}+\frac{2}{M_0+t_0}}{M_0+t_0}e^{-Lt_0 /\hbar v}.\nonumber
\end{eqnarray}\noindent So initially if the
in-plane angle difference of two magnets is zero, i.e. $(\theta_L=\theta_R=0)$, 
the two boundstates have in-plane magnetic moments aligned and anti-aligned with the magnetization, respectively (as shown in Fig. \ref{fig:energySJE}).
 Now if we  fix $\theta_L=0$ and set $\theta_R=\Delta\theta$  then the
magnetic moment of the bound state is frustrated in that it encounters an ambiguity in the direction. The optimum choice is for the boundstates to pick a compromising direction between the two external moments, specifically,  $\frac{\theta_{L}+\theta_{R}}{2}=\Delta\theta/2.$  Hence, rotating $\theta_R$ by $2\pi$ causes the magnetic moment of the boundstates to only rotate by  $\pi.$ The state that was initially aligned becomes anti-aligned (and vice-versa), and in order to return to the initial state $\theta_R$ must rotate by an additional $2\pi.$

The nature of the $4\pi$ periodicity can be further gleaned by formally decomposing the boundstate operators at the two interfaces $b_{L}, b_{R}$ into pairs of Majorana fermions $b_{L}=(\eta_1+i\eta_2)/2,\;\;\; b_{R}=(\gamma_1+i\gamma_2)/2.$  In the basis of Majorana operators the effective boundstate Hamiltonian can be expressed as two separate copies of the effective Majorana Hamiltonian in the fractional charge Josephson effect\cite{FuKaneJJ}, {\it i.e.}, $H(\Delta\theta)=(i/2)F(\Delta\theta)\left(\gamma_1\eta_1+\gamma_2\eta_2\right)$ where $F(\Delta\theta)$ is the $4\pi$-periodic function defined in Eq. \ref{eq:Ftheta}. In terms of  new complex fermion operators $d_1=(\gamma_1+i\eta_1)/2,\; d_2=(\gamma_2-i\eta_2)/2$ which are combinations of \emph{both} $b_{L/R}$ and $b^{\dagger}_{L/R}$, the Hamiltonian is simplified to $H(\Delta\theta)=F(\Delta\theta)\left(d^{\dagger}_{1}d^{\phantom{\dagger}}_1-d^{\dagger}_{2}d^{\phantom{\dagger}}_{2}\right)$ and can be thought of as a pseudo-spin degree of freedom in a $\Delta\theta$-dependent Zeeman field.  When $\Delta\theta\to\Delta\theta+2\pi$ the lowest energy pseudo-spin state flips direction and only returns back to the initial state when the phase-difference advances by $4\pi,$ as expected. 
For example, for the case $\theta_L=0,\theta_R=\Delta\theta=2\pi$ we have seen that  $b_{L}\to b_{L}$ and $b_{R}\to -b_{R}$ which means $\gamma_{1/2}\to-\gamma_{1/2}.$ This implies that this shift of $\Delta\theta$ sends $d_{1/2}\to d_{1/2}^{\dagger}$ which switches leaves the fermion parity invariant since it transforms \emph{both} particle states to hole states.This physics is reminiscent of the fermion parity flip seen in the fractional charge Josephson effect\cite{FuKaneJJ}. Here, while the doubling of the Majorana fermions renders the parity a constant, the pseudo-spin of the lowest energy bound state flips for every advance of $2\pi.$

Turning now to observation of the SJE,  measuring the spin current, perhaps through magnetic or optical means, would be an obvious possibility but an electrical detection method would be experimentally preferable to other detection schemes. As a corollary, we find that as $\Delta\theta$ changes the charge density between the ferromagnets oscillates between the two edges with a $4\pi$ periodicity. The probability that the low-energy boundstates lie on the top or bottom edges behaves, for the boundstate with $E(\Delta\theta)=\pm F(\Delta\theta)$, as
\begin{eqnarray}
P_{top}=\frac{1}{2}\pm \frac{\Xi}{2}\sin\frac{\Delta\theta}{2},\;\; P_{bottom}=\frac{1}{2}\mp \frac{\Xi}{2}\sin\frac{\Delta\theta}{2}.
\end{eqnarray}\noindent  Thus the voltage drop between points A and B in Fig. \ref{twoedge} would also show a $4\pi$ periodic signal. While the voltage signal would be small, since it is essentially coming from the fluctuation of a single charge, it should be possible to measure using single-electron transistor/Coulomb blockade techniques (see \emph{e.g.} Ref. \cite{yacoby2004}). 

A dual effect can be induced by applying a voltage difference between the two edges, as captured by $H_V=\frac{1}{2}V\tau^z.$ The bound state energies and spin-current become
\begin{eqnarray}
E_{\pm}&=&\pm \left(\Gamma\cos \tfrac{\Delta\theta}{2}+\frac{1}{2}V\Xi \sin \tfrac{\Delta\theta}{2}\right)=\pm J_S\cos\tfrac{\Delta\theta-\phi_0}{2}\nonumber\\ \\
I_s&=&\pm\tfrac{J_S}{2}\sin\tfrac{\Delta\theta-\phi_0}{2}
\end{eqnarray}\noindent where $J_S=\sqrt{\Gamma^2+V^2\Xi^2/4}$ and $\phi_0=2\arctan \frac{V\Xi}{2\Gamma}.$ Thus the spin current can be adjusted by applying an inter-edge voltage difference (as seen in the  $\phi_0$ dependence). For example, even if $\Delta\theta=0$ we can turn on the spin-current by applying a voltage and as $V\to \infty$ we see that the spin-current reaches a maximum as $\sin(\phi_0/2)\to 1.$ This physical phenomenon is like the intrinsic spin-Hall effect where the an applied voltage generates a  spin-current flowing perpendicular to the electric field. Thus, as indicated in our earlier arguments, the spin-current induced from a $\Delta\theta$ will produce an inter-edge voltage due to an inverse spin Hall effect. Moreover, this voltage term generates a spin-Josephson $\phi_0$ junction which is an analog of the Josephson $\phi_0$ junction\cite{Buzdin2008}.

In analogy with the SC case, we also consider the AC SJE in the presence of an inter-edge voltage. The effective low-energy Hamiltonian of FM/TI/FM junction can be written as:
\begin{equation}
H_{SJ}=-J_S\cos\tfrac{\Delta\theta-\phi_0}{2}+\alpha(S^{z}_{R})^{2}+\alpha (S^{z}_{L})^2,
\end{equation}
where $\alpha$ represents the easy-plane anisotropy energy. 
 Using the canonical relations introduced in Eq. \ref{FMcommutation}
we can derive the Josephson relations
\begin{equation}
I_s=-\tfrac{J_S}{2}\sin\tfrac{\Delta\theta-\phi_0}{2},\;\;\; \Delta{\dot{\theta}}=2\alpha (S_{R}^{z}-S_{L}^{z})
\end{equation}
Thus an $S^z$ imbalance acts like a `spin-voltage' and results in a time-dependent $\Delta\theta.$ If one induces a static $S^z$ imbalance using applied magnetic fields then there would be an AC fractional SJE current. In addition, an oscillating voltage signal would be present from the same mechanism as  the above which can be measured using voltage probes or via the accompanying microwave radiation\cite{KwonSenguptaYakovenko}.

Finally, we discuss two issues for the measurement of the fractional SJE, first, a stringent  requirement of particle-hole symmetry. This naturally appears in the SC case because of the Bogoliubov-de Gennes redundancy, but is absent in the SJE setting. The presence of a local potential would therefore move bound states away from zero energy, obscuring and altering the spin-Josephson signal. 
To fix this issue we require that a gate (which applies the same voltage to both edges) be available to locally tune the boundstate energies in the tunnel-junction region. The other consideration is that in the real 2D TI material there exist inversion symmetry breaking terms that remove the conservation of the $S_z$ spin carried by the spin-Josephson current. Since our prediction does not involve a quantized signal, the primary effect of the (usually very weak) non-conservation terms would be to reduce the amplitude of the spin-current from our calculated value. Thus, assuming that we can add a gate to our system, neither of these issues qualitatively alter our predictions.

We have thus shown that several testable effects appear in ferromagnetic tunnel junctions which are unique to 2D topological insulators. 
In conjunction with the work of some of the authors in Ref. \onlinecite{foldingpaper}, a number of predictions have now been made based on the understanding that the magnetic gap and the tunneling gaps \emph{compete} and can yield bound states. The advantage of having ways to electrically perturb and measure these spin effects makes experimental observation more accessible .
We are optimistic that these rich phenomena are robust, devoid of ultra-fine tuning, and can be observed in systems such as HgTe/CdTe quantum wells. 

\emph{Note:} During the preparation of this manuscript we became aware of overlapping work by  L. Jiang, D. Pekker, J. Alicea, G. Refael, Y. Oreg, A.Brataas and F. v. Oppen in arxiv: 1206.1586. Additionally a preprint with some overlapping material recently appeared (arxiv: 1206.0776).

\emph{Acknowledgements}
 This work is supported by the U.S. Department of Energy, Division of Materials Sciences under Award No. DE-FG02-07ER46453 (Q.M., V.S., T.L.H. and S.V) and the AFOSR under grant FA9550-10-1-0459 (MJG) and the ONR under grant N0014-11-1-0728 (MJG)

%

\end{document}